# Trapping/Pinning of colloidal microspheres over glass substrate using surface features

Praneet Prakash[1], Manoj Varma[1,2]

[1]Centre for Nano Science and Engineering, Indian Institute of Science, Bangalore

[2]Robert Bosch Centre for Cyber Physical Systems, Indian Institute of Science, Bangalore

[*]mvarma@iisc.ac.in

## Abstract

Suspensions of micro/nano particles made of Polystyrene, Poly(methyl methacrylate), Silicon dioxide etc. have been a standard model system to understand colloidal physics. These systems have proved useful insights into phenomena such as self-assembly. Colloidal model systems are also extensively used to simulate many condensed matter phenomena such as dynamics in a quenched disordered system and glass transition. A precise control of particles using optical or holographic tweezers is essential for such studies. However, studies of collective phenomena such as jamming and flocking behaviour in a disordered space are limited due to the low throughput of the optical trapping techniques. In this article, we present a technique where we trap and pin polystyrene microspheres ~ $10\,\mu m$ over 'triangular crest' shaped microstructures in a microfluidic environment. Trapping/Pinning occurs due to the combined effect of hydrodynamic interaction and non-specific adhesion forces. This method allows trapping and pinning of microspheres in any arbitrary pattern with a high degree of spatial accuracy which can be useful in studying fundamentals of various collective phenomena as well as in applications such as bead detachment assay based biosensors.

## Introduction

Understanding of collective phenomena such as agglomeration, jamming, glass transition, directed self-assembly is a major research area in mesoscale physics. The underlying physics behind these phenomena occur at a length scale intermediate between atomic and bulk matter. Typically an aqueous solution of micro/nano sized polystyrene or silica particles is used in experimental studies. They interact weakly among themselves through van der Waals, Coulomb and depletion forces. These interactions can be easily tuned by functionalizing the particles[1–3] or changing their chemical environment.[4,5] Therefore, colloidal solutions can be used to study many condensed matter phenomena. In particular, colloidal microspheres have been an excellent model system to gain an understanding of long-standing open issues associated with phase transitions in glassy systems, for instance, the role of quenched disorder in glass transition.[6–8] In general, the role of disorder in a variety of collective phenomena has attracted recent interest. For instance, researchers are studying collective behaviour of active particles in disordered media motivated by the desire to understand motility of living organisms such as bacteria in real environments.[9–11] A relatively simple way to create disordered energy landscapes to study such phenomena is by trapping or pinning particles at desired spatial locations. Conventionally, an optical tweezer is used to trap and manoeuvre such particles.[12,13] Recently, researchers have shown trapping of microspheres in microfluidic environment by various strategies such as the creation of eddies using oscillating flow around obstruction[14], fluid flow from multiple directions to create low pressure region[15], use of rotating magnetic rod to create vortex[16]. Though these systems are very useful, they have certain limitations. The applicability of an optical tweezer is limited by the laser intensity which a sample can withstand. Trapping methods based on hydrodynamics require very specific settings in each case, and hence lacks versatility. In general, the trapping methods developed so far have a low throughput, which fundamentally restricts the study of collective phenomena. To address the issue of throughput, modification such as holographic optical tweezers have been used to create pinning patterns[17,18] but, the number of optical traps which can be generated is limited, typically of the order of ~ 100.[19] The techniques developed as yet are largely non-contact, presumably to provide a pristine system for experiments. However, a non-contact mode of trapping may not be a critical requirement for many experiments in colloidal physics. The contact mode of trapping can be done by flowing functionalized particles over a chemically patterned substrate where particles get pinned due to the formation of a chemical bond.[20–22] A contact mode of trapping such as the one mentioned above requires multiple steps of chemical patterning which makes the process tedious and less versatile. However, such methods enable the creation of well defined pinning patterns of particles which are necessary to understand the effect of disorder in collective behavior. The complexity of chemical approaches to create pinning patterns prompted us to consider a purely physical approach towards contact-mode of trapping (In this article, we use the terms trapping and pinning synonymously). Specifically, we explored the ability of microstructured surface[23-25] features to trap/pin microspheres using static potential wells generated by near-field flow effects from microstructures inside a microfluidic channel. We demonstrate, in this article, a simple process to create an arbitrary 2D pattern of pinned particles using microstructures whose height could be as low as 1/10th of that of the particles.

The adhesion force between $5 - 10\ \mu m$ polystyrene microspheres and the glass substrate is of the order of $1 - 100\ nN$.[26,27] In comparison, force due to gravity on the microspheres in aqueous medium is less than $1\ pN$. At such range of adhesion strengths, microspheres can be reliably pinned if they come in contact with the substrate surface. The flow of an aqueous solution over a microstructured substrate having a shape of triangular crest in a microfluidic channel leads to a formation of static low pressure region around it (Fig. 1(d)). Microsphere laden flow over such a substrate in a microfluidic channel leads to particles getting pinned in the low pressure region around the triangular crests. We found such structures to be very conducive for reliable trapping/pinning of microspheres of diameter as low as $5\ \mu m$. We have shown, robust trapping/pinning over a crest of height as low as $\sim 1\ \mu m$, in comparison, the microsphere diameter was $5\ \mu m\ \&\ 10\ \mu m$. We further explored the critical role of microstructure geometry and flow rate in reliable trapping/pinning of particles which is supported by fluid dynamics simulation using the finite element method (FEM) implemented in COMSOL, a commercial numerical package. The technique described here offers a simple method to trap and pin polystyrene particles in 2D patterns with minimal complexity and can be easily scaled up.

## Materials and Methods

### Channel design and trapping principle

A schematic of the microfluidic channel used for the experiments is shown in Fig. 1(a). A PDMS channel is clamped onto an etched glass slide which was fabricated by an isotropic hydrofluoric acid (HF) etch of a photoresist patterned substrate. The channel has two inlets, one for the microsphere solution and the other for the Phosphate Buffered Saline (PBS buffer) solution to generate shear force by means of the fluid flow. Polystyrene microspheres sized $5\ \mu m\ \&\ 10\ \mu m$ in $0.01M$ PBS buffer solution were flown through the microsphere solution inlet. As they flow past near the crests they get pinned in the low pressure region (Fig. 1(d)) by adhering to the substrate as depicted in Fig. 1(b), 1(c). The probability of a particle getting trapped in the upslope or downslope regions depends on the flow rate. At high flow rates, particles are predominantly pinned in the downslope region while at a very low flow rate pinning can be also seen in upslope region.

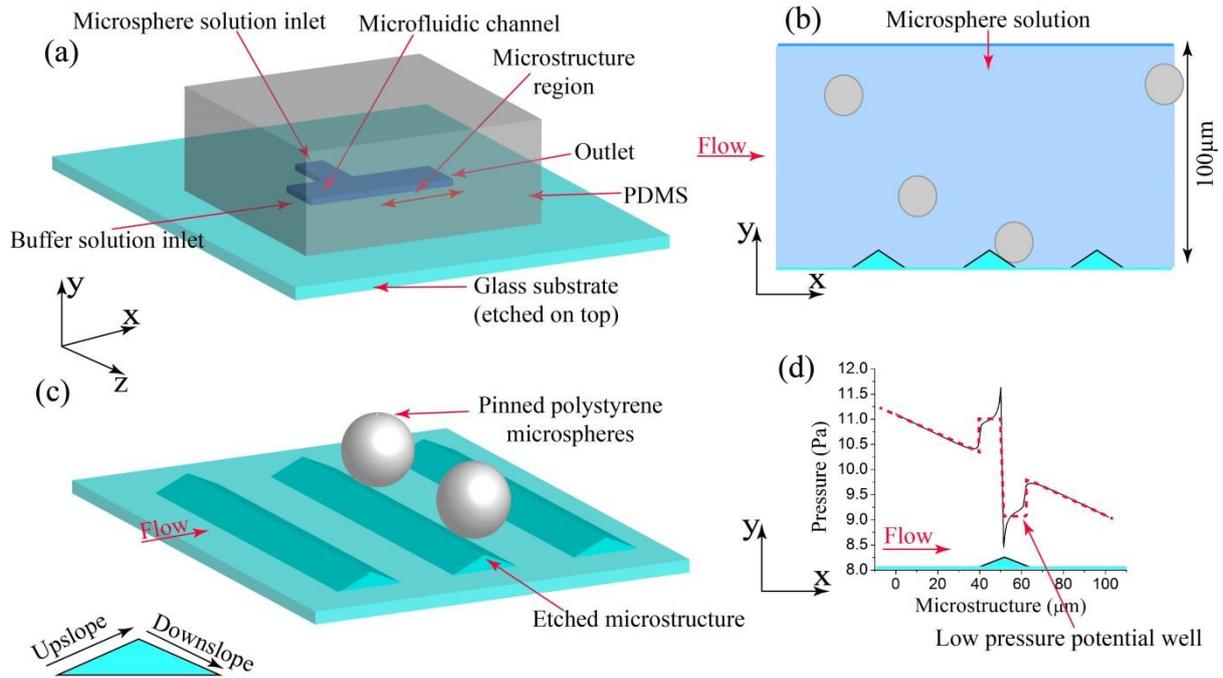

**Figure 1** Schematic of a PDMS microfluidic channel clamped over an etched glass substrate. (a) T-shaped microfluidic channel with two inlets, one for microsphere solution and other for the plain buffer solution. (b) Side view of microfluidic channel showing pinned polystyrene microspheres. Channel has a width of $500\ \mu m\ (z-axis)$ and a depth of $100\ \mu m\ (y-axis)$, length of the entire channel is $1\ cm\ (x-axis)$. (c) Top view of the pinned polystyrene microspheres over etched microstructure. (d) The simulated pressure profile over the micro-structured glass substrate of height $4\ \mu m$ and width $20\ \mu m$ at the bottom of the channel. It was calculated by setting up the Navier-Stokes equation for an incompressible fluid flow at a flow rate of $50\ \mu l/min$.

## Experimental setup

A T-shaped Polydimethylsiloxane (PDMS) microfluidic device with a channel depth of $100\ \mu m$, width $500\ \mu m$ and length $1\ cm$ was cured over a SU8 mould fabricated using photolithography process. The microfluidic device was assembled by clamping PDMS channel on the microstructured glass substrate using binder clips (Fig. 2), and observed using an inverted imaging setup where the device was illuminated from the top and imaged from the bottom.

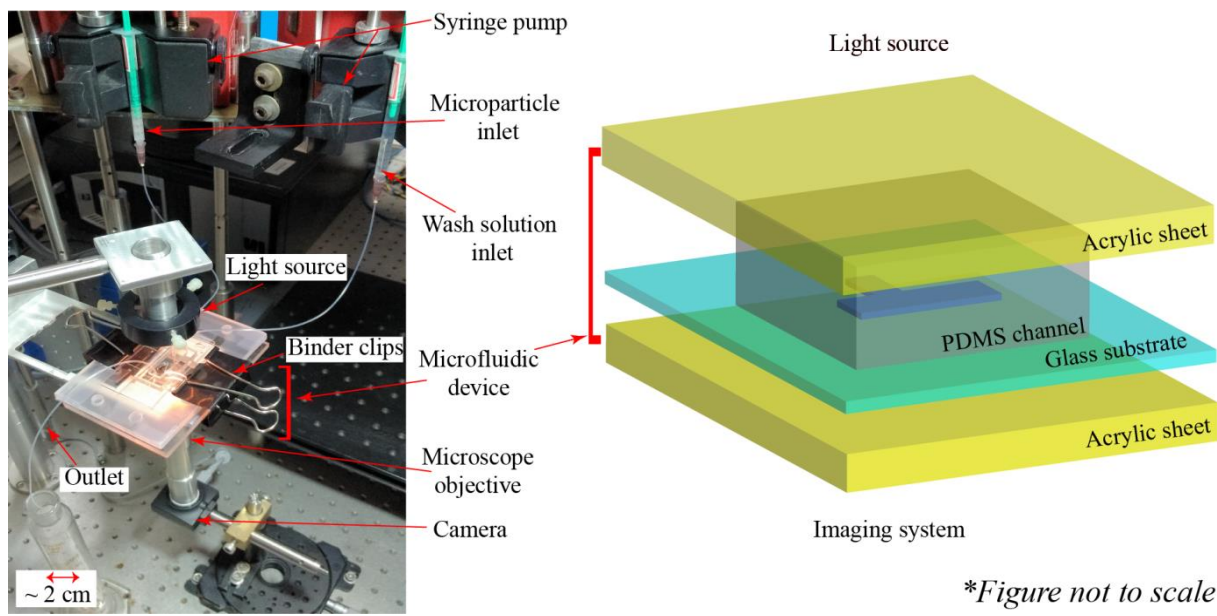

**Figure 2** Schematic of the experimental setup: A microfluidic channel made up of PDMS and glass substrate is sandwiched between the top and bottom layer of acrylic sheet and the whole assembly is clamped tightly using binder clips. Device is illuminated from the top and imaged from the bottom using $20x\ \&\ 40x$ microscope objective along with a CMOS camera assembly.

**Fabrication of the microfluidic channel**

A $100\ \mu m$ thick layer of negative resist SU-8 2100 was spin coated on a silicon wafer and then patterned using photolithography technique. Subsequently, it was developed to achieve a $500\ \mu m$ wide, $100\ \mu m$ thick and $1\ cm$ long T-shape mould. The channel was fabricated by pouring PDMS and curing agent in the ratio of $10:1$ over the mould and was cured at a temperature of $100°C$ for an hour. After curing, the microfluidic channel was peeled off and the inlet/outlet holes were punched using a biopsy punch.

**Fabrication of microstructure**

The role of microstructure in trapping and pinning of polystyrene microsphere is determined by performing the experiments on microstructures of various heights and widths (Fig. 3). Commonly available microscopic glass slides of dimension $75\ \text{mm} \times 25\ \text{mm}$ and thickness of $1\ mm$ were used as a substrate. Resist patterned glass slides were isotropically etched using a mixture of $50\%\ \text{HF}, 36\%\ \text{HCl}$ and Deionized (DI) water in the ratio $1:1:3$. HF etches out various silicon oxides present in the glass slide while HCl removes metal contents sodium and calcium respectively. This mixture gives an etch rate of about $100\ \text{nm/s}$. As wet etching is an isotropic process, it etches out the glass substrate in transverse as well as lateral direction. Lateral etching was controlled to some extent by optimising the post-development baking time of photoresist. We used positive photoresist S1813 for microstructures with depth $1 - 2\ \mu m$ and AZ4562 for even larger depths which resulted in an optimal etching and resist stripping process. Fig. 3 shows the Atomic Force Microscopy (AFM) micrograph of three microstructures with different dimensions used for the experiments.

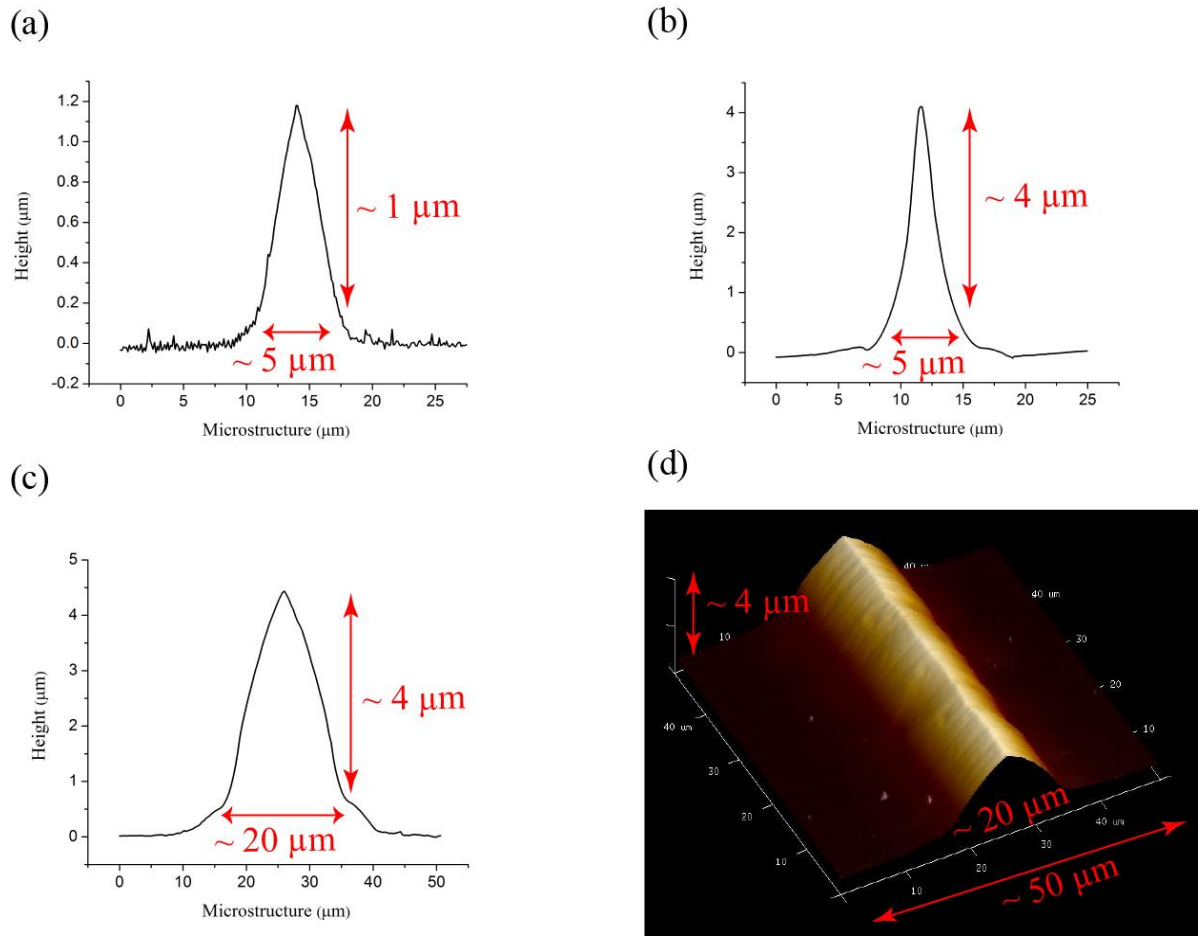

**Figure 3** AFM micrograph of the microstructures. Microstructure dimension: $height\ (h)$, $width\ (w)$. (a) $h = 1.1\ \mu m$, $w = 5.1\ \mu m$. (b) $h = 4.1\ \mu m$, $w = 5.4\ \mu m$. (c) $h = 4.3\ \mu m$, $w = 19.2\ \mu m$. (d) Three-dimensional AFM micrograph of the microstructure of $h = 4.3\ \mu m$, $w = 19.2\ \mu m$.

The etch time and chemicals used for fabricating microstructures are mentioned in supplementary Table 1. We fabricated an array of the microstructures of different dimensions as shown in Fig. 3(a), 3(b), 3(c) with a periodicity of $50\ \mu m, 50\ \mu m$ & $100\ \mu m$ respectively.

The microstructure of smaller height, as shown in Fig. 3(a), was fabricated by using photoresist S1813, it was coated @ 2000 rpm over glass substrate which provides a uniform resist thickness of about $2.2\ \mu m$. Subsequently, the resist was photo-lithographically patterned to get an array of width $5 \mu m$ and periodicity $50\ \mu m$ over the substrate. A post-development bake of $3\ min$ @ 110°C was done before dipping it in the HF etchant for

$10\ sec$ to etch out the bare glass portion. Similarly for the triangular crest structure of larger dimensions as shown in Fig. 3(b) & 3(c), the substrate was coated with AZ4562 @ 4000 rpm providing a resist thickness of about $6.2\ \mu m$ which was then patterned to get a resist array of width $5\ \mu m$ & $20\ \mu m$ respectively. A post-development bake of $5\ min$ @ $110°C$ was done before dipping it in the etchant solution for 40 sec to get an array of microstructures. The residual resist was removed by the sonication of substrate in acetone. The microstructured glass substrate was further cleaned in piranha solution ($3\ H_2SO_4 + 1\ H_2O_2$) before the experiment. A roughness measurement over 5 glass substrate using Bruker AFM tapping mode revealed little change post etching. The average roughness pre and post etching where about $7.8\ nm$ and $8.3\ nm$ respectively.

**Chemicals and microsphere solution**

$10\mu m$ & $5\mu m$ Polystyrene microspheres were purchased from Sigma-Aldrich. They were rinsed using DI water and diluted to $10\ mg$ per $ml$ using Phosphate Buffered Saline (PBS) $0.01M$. PBS buffer was prepared by mixing PBS tablets from Sigma-Aldrich in $DI$ water and has a pH of 7.4 at $25°C$. The number of $10\mu m$ & $5\mu m$ particles in solution is of the order of $10^6$ & $10^7$ per $ml$ respectively. $0.1\ ml$ of microsphere solution was used in a single experiment.

**Numerical simulation of pressure profile over crests**

The pressure profile in the channel was calculated by solving the Navier-Stokes equation (Equation 1) corresponding to incompressible flow for which $d\rho/dt = 0$, which is equivalent to Equation 2.

$Navier-Stokes\ Equation$: $\quad\quad\quad \rho\frac{\partial u(t)}{\partial t} + \rho(u \cdot \nabla u) = -\nabla p + \mu \nabla^2 u \quad\quad\quad (1)$

$Continuity\ Equation$: $\quad\quad\quad\quad\quad\quad\quad \nabla \cdot u = 0 \quad\quad\quad\quad\quad\quad\quad (2)$

where $\rho$ & $\mu$ is the density and dynamic viscosity of the solution. A two-dimensional schematic of the channel is shown in Fig. 1(c), where, no-slip boundary condition ($u = 0$) is applied at the top and bottom surface along with an inflow and outflow condition depending upon the flow rate ($u = u_{avg}$) at the inlet and outlet of the channel respectively. Fig. 4 shows the simulated pressure profile arising from fluid flow across a microstructure. A low pressure potential well can be seen in the upslope and downslope region of the triangular crest. At a very low flow rate polystyrene microspheres get pinned in the upslope region simply because the microstructure acts as a barrier to their flow. However, at a higher flow rate, polystyrene microspheres accelerate as they flow through the upslope region due to decrease in cross-sectional channel volume. As a result they acquire a velocity in $+y$ direction thus moving away from the substrate resulting in no pinning. On the contrary, in the downslope region microspheres decelerate due to increase in the channel volume and acquire a velocity in $-y$ direction thus moving towards the substrate which results in an increased probability of pinning. Surface plots of the velocity vector are shown in supplementary Fig. S1.

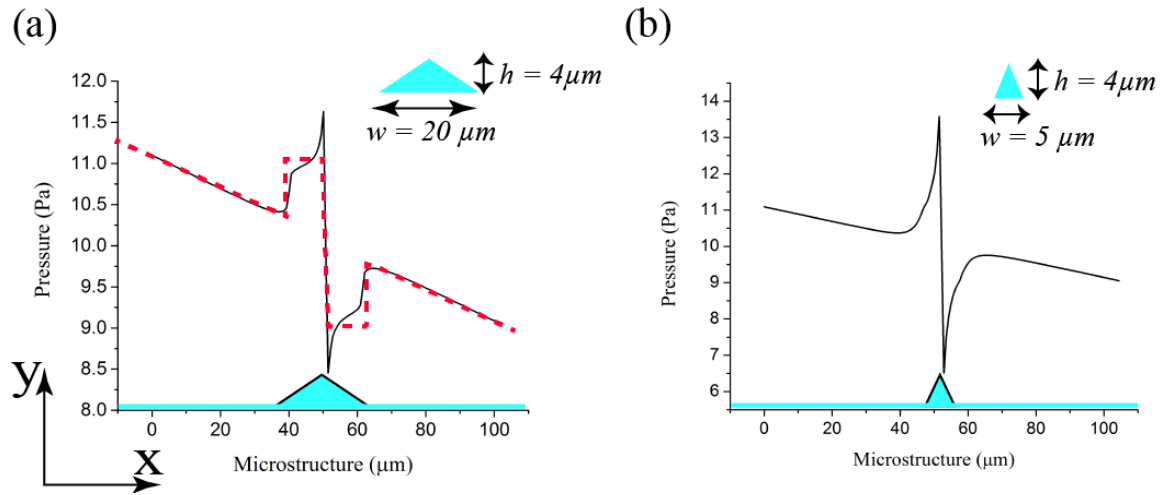

**Figure 4** Pressure profile plots over the surface of microstructure with different geometries showing low pressure region around crest. Volumetric flow rate was $50 \mu l/min$ and the channel domain was considered to be $1\ mm$ long instead of $1\ cm$ to save the computation cost. Microstructure dimension: $height\ "h"$, $slope\ width\ (upslope + downslope\ region - "w")$ (a) $h = 4\ \mu m, w = 20\ \mu m$. (b) $h = 4\ \mu m, w = 5\ \mu m$.

Pressure profile over microstructures of different shapes is shown in Supplementary Fig. S2. We focussed on the 'triangular shaped crests' as they can be easily fabricated using isotropic etch of glass substrate.

**Estimation of adhesion force between microsphere and glass substrate**

Non-specific forces such as the van der Waals and electrostatic double layer interaction force, collectively known as the DLVO force, exist between any two surfaces. At macro-scale, their effect is minuscule and not readily observable. However, they can be prominent at micro scale. Microsphere detachment experiments using fluidic shear force was done on plain glass substrates to estimate the adhesion strength.[27–29] Polystyrene microspheres of size $5 \mu m\ \&\ 10 \mu m$ were flown along with buffer solution through the microsphere inlet. After sustaining a stable flow, the flow was stopped so that the microspheres can settle on the bottom of the substrate. Velocity in the downward direction due to the gravitational force of a

10 $\mu m$ sized microsphere is around 2.75 $\mu m/s$ and therefore it would take about a minute for all the microspheres to settle. The 5 $\mu m$ particles have a downward velocity of about 0.69 $\mu m/s$ and take a longer time to settle. After a 5 minute of incubation period, PBS buffer solution was flown through the buffer inlet and the flow rate was increased until the viscous drag was enough to detach the particles. We saw no detachment up to a flow rate of 500 $\mu l/min$ and 300 $\mu l/min$ for microspheres sized 5 $\mu m$ & 10 $\mu m$ respectively in the experiment done on five different glass substrates. The microsphere is assumed to be static and the no slip boundary condition is applied on the channel surface as well as on the microsphere to estimate the shear force using simulation, as described in previous section. The viscous drag on microspheres sized 5 $\mu m$ & 10 $\mu m$ adhered to the bottom surface of the channel comes out to be 1.88 $nN$ & 4.6 $nN$ respectively which matches well with the analytical estimation explained in supplementary text. The larger drag force on the microsphere sized 10 $\mu m$ even at a lower flow rate is due to the parabolic flow profile in a rectangular channel. The flow rate in the entire experimental analysis is less than 100 $\mu l/min$, well within the upper limit of microsphere detachment flow rate of 300 $\mu l/min$.

## Results and Discussion

The density of polystyrene is 1.05 $gm/cm^3$ while that of water is 1.0 $gm/cm^3$. Hence, in an aqueous medium, they move downward due to the gravitational force. If these particles are pumped in a microfluidic channel over a plain glass substrate and left idle for some time they will randomly settle within the channel and can be reliably pinned. However, if they are pumped continuously they won't pin at all. Even at a low flow rate of 5 $\mu l/min$, the microspheres simply glide over the substrate and are flushed out. We have used an array of microstructures to demonstrate high throughput trapping/pinning of polystyrene microspheres. Fig. 5(a) & 5(b) shows pinned microspheres of diameter 10 $\mu m$ & 5 $\mu m$ respectively on the microstructures of smaller dimension ($h = 1.1\ \mu m, w = 5.1\ \mu m$). The microspheres typically pin at flow rates less than 50 $\mu l/min$, pinned microspheres as shown in Fig. 5(a), 5(b) and 5(c) is achieved by stopping the flow for $30 - 40\ sec$ at a high flow rate of 50 $\mu l/min$ which results in a high throughput. A real-time video of trapping/pinning of particles is shown in the movie "$supple\_pinning1.mov$" provided as a supplementary data.

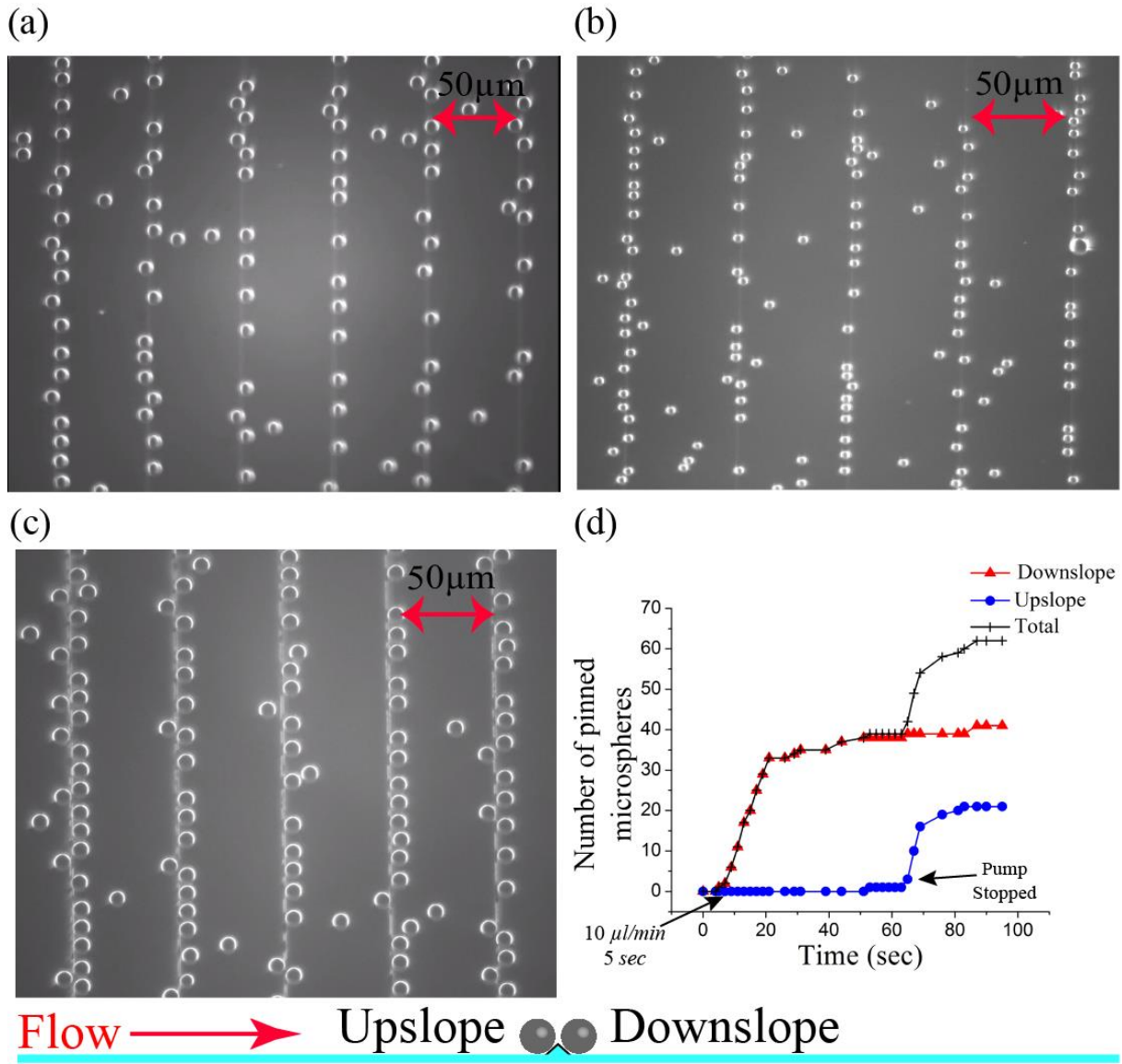

**Figure 5** Images of pinned microspheres on an array of microstructures. (a) Microspheres of diameter 10 $\mu m$ pinned on the microstructure of $h = 1.1\ \mu m$, $w = 5.1\ \mu m$. (Fig. 3a). (b) Microspheres of diameter 5 $\mu m$ pinned on the microstructure of $h = 1.1\ \mu m$, $w = 5.1\ \mu m$ (Fig. 3a). (c) Microspheres of diameter 10 $\mu m$ pinned on the microstructure of $h = 4.1\ \mu m$, $w = 5.4\ \mu m$. (d) A plot of number of pinned microsphere vs time in the upslope and downslope region cumulated over five microstructure of $h = 1.1\ \mu m$, $w = 5.1\ \mu m$.

Fig. 5(d) shows the number of pinned microsphere as the time progresses. Initially, the flow rate is 50 $\mu l/min$ which is reduced to 10 $\mu l/min$ at $t = 5\ sec$ resulting in the pinning ar the downslope region which saturates after some time due to a decrease in the number of empty sites. Later, the pump is stopped at $t = 60\ sec$, as the speed further decreases microspheres starts filling the upslope region. Microsphere sized $10 \mu m$ has a velocity of $\sim 2.75\ \mu m/s$ in the downward direction, at this velocity it takes roughly $0.35\ sec$ for the particles to cover a distance of $1 \mu m$ which is the height of microstructure shown in Fig. 5a & 5(b). The periodicity of the microstructure is $50\ \mu m$, hence, velocity of microspheres sized $10\ \mu m$ along the channel should be 145 $\mu m/s$ (volumetric flow rate = 18 $\mu l$/min), such that it encounters the successive crest before touching the substrate surface. This condition gives a lower limit of the flow rate for a precise pinning in the downslope region.

**Pinning efficiency of the microstructured substrate**

Fig. 6(a) and 6(b) shows the histogram of pinned microsphere of size 10 $\mu m$ cumulated over six microstructures which we were able to image at once. The length of each microstructure is 360 $\mu m$, hence, at the most 36 microspheres can be pinned along each microstructure. If we assume that on an average pinned microspheres are equally spaced then roughly half of the sites can't be occupied and hence a total of $18 \times 6 = 96$ sites are available. From the histograms in Fig. 6(a) & 6(b) we can see that up to 100% of the sites can be filled within two minutes under appropriate conditions. Furthermore, as the height is increased from 1.1 $\mu m$ in Fig. 6(a) to 4.1 $\mu m$ in Fig. 6(b) microspheres can be pinned even at a higher flow rate.

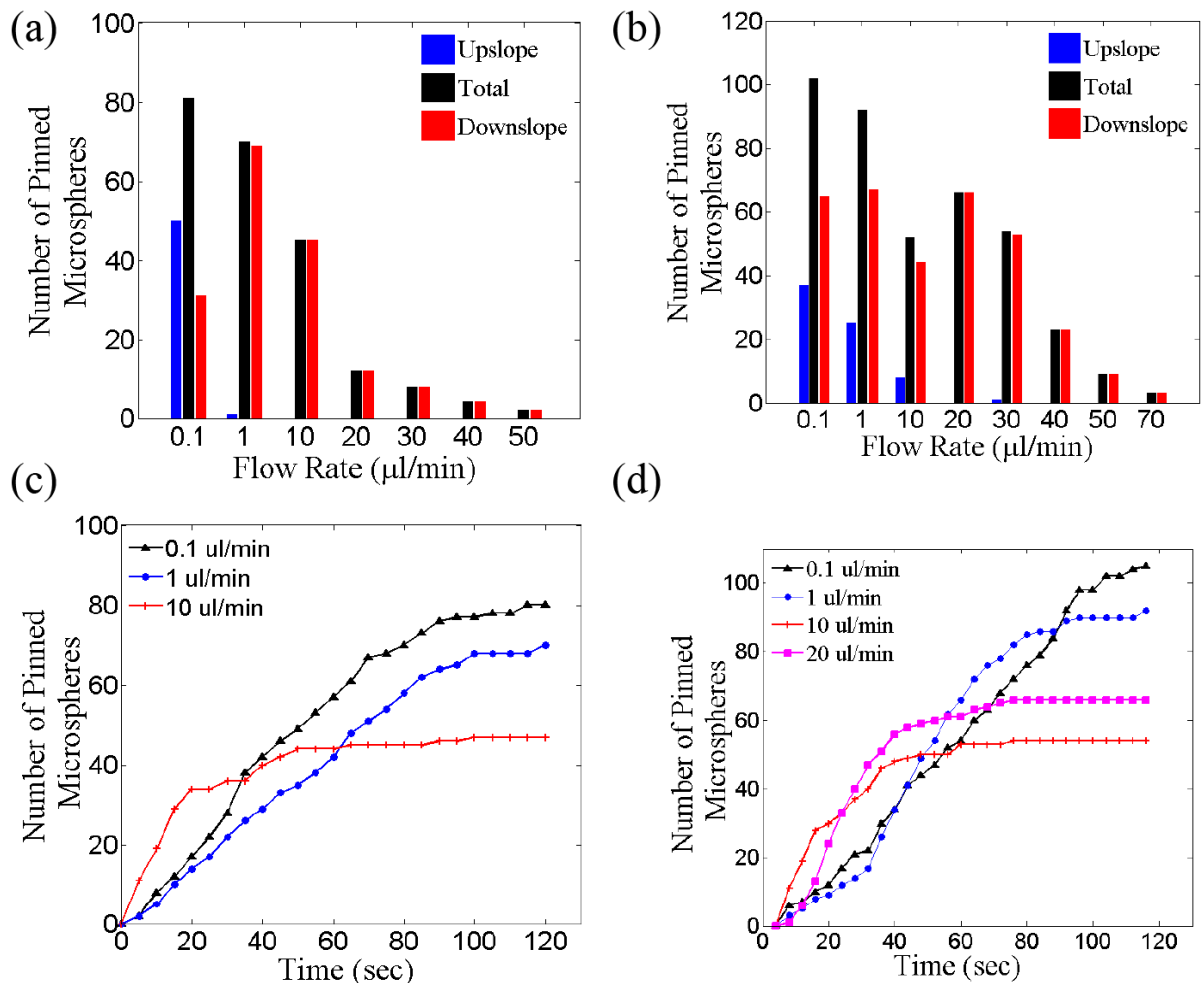

**Figure 6** Histogram of the number of pinned microspheres at different flow rates in 2 minute for microstructure

(a) $h = 1.1\ \mu m,\ w = 5.1\ \mu m$ ; (b) $h = 4.1\ \mu m,\ w = 5.4\ \mu m$. Graph of number of pinned microspheres vs time at different flow rate for microstructure (c) $h = 1.1\ \mu m,\ w = 5.1\ \mu m$ ; (d) $h = 4.1\ \mu m,\ w = 5.4\ \mu m$.

Fig. 6(c) and 6(d) shows the number of pinned microspheres with time at different flow rates. The number of pinned microspheres rises linearly and persists with time for flow rates below $1\ \mu m$ whereas it saturates quickly after increasing linearly at higher flow rates $\geq 10\ \mu l/min$. The microspheres pin at low flow rates $\leq 1\ \mu l/min$ in upslope region because the crests act as a gravitational potential barrier. A real time video "$supple\_pinning2.mov$" is provided which shows pinning of microsphere in the upslope region $(t = 0 - 15\ sec)$ at a low flow rate of $0.1\ \mu l/min$ on the microstructure of $h = 1.1\ \mu m,\ w = 5.1\ \mu m$. Velocity of the microspheres which were pinning in the upslope region was in the range of $6 - 18\ \mu m/s$ as estimated from the video analysis.

**Pinning of microspheres at a high flow rate over larger microstructure**

Robust pinning at a high flow rate of $50\ \mu l/min$ is achieved on the larger microstructures of $h = 4.3\ \mu m, w = 19.2\ \mu m$ as shown in Fig. 7(a) & 7(b). The drag force on the pinned microspheres as depicted in Fig. 7(d) is $\sim 0.83\ nN$ which is much lower than the upper limit of shear detachment force $\sim 4.6\ nN$ at a flow rate of $300\ \mu l/min$.

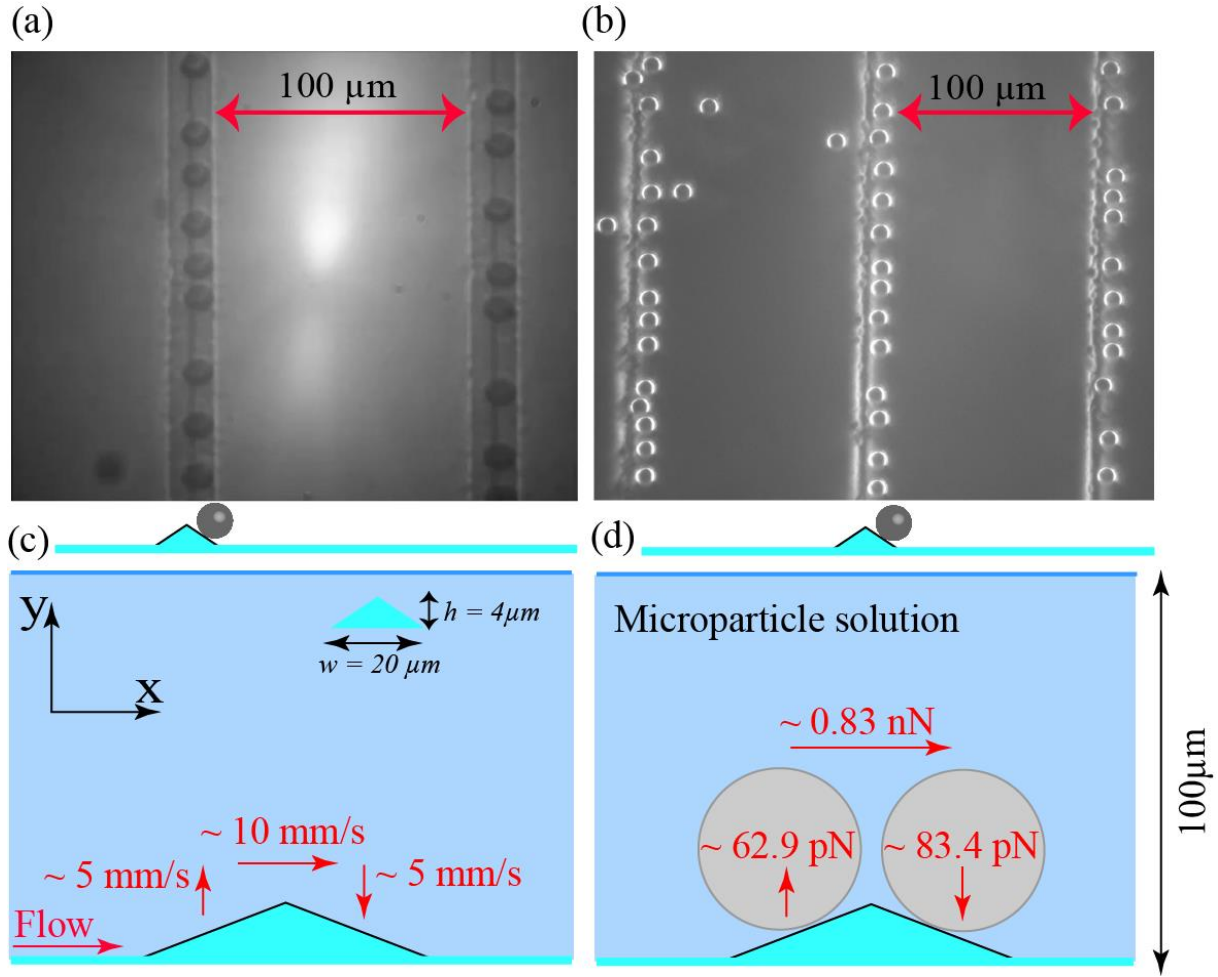

**Figure 7** Pinning at high flow rates over the microstructure of $h = 4.3\ \mu m, w = 19.2\ \mu m$. (a) Bright field expanded image of pinned microspheres of size $10\ \mu m$ on the downslope region of microstructure $h = 4.3\ \mu m$, $w = 19.2\ \mu m$. (b) Dark field image of $10\ \mu m$ microspheres preferentially pinned on the downslope region of microstructure $h = 4.3\ \mu m, w = 19.2\ \mu m$. (c) Average velocity along the channel just above the microstructure is $10\ mm/s$ whereas in upslope and downslope region, it is $5\ mm/s$ in $(+y)$ and $(-y)$ direction respectively. (d) Microsphere of size $10\ \mu m$ pinned in upslope region experiences a lift force of $62.9\ pN$ thus opposing the adhesion whereas in the downslope region, a force of $83.4\ pN$ is experienced in the $(-y)$ direction, hence assisting the adhesion. Drag force along the flow at both upslope and downslope region is nearly $0.83\ nN$.

We can see from the histogram in the Fig. 6(b) that very few microspheres are pinned at a flow rate of $50\ \mu l/min$ on the microstructure of $h = 4.1\ \mu m, w = 5.4\ \mu m$. However, considerable pinning is seen as the width of the microstructure is increased from $5.4\ \mu m$ to $19.2\ \mu m$, as shown in Fig. 7(a) & 7(b). Hence, trapping/pinning can be achieved at a high flow rate either by increasing the height or the width of the microstructure. A microsphere on the top of the microstructure flows with an average velocity of $10\ mm/s$ as depicted in Fig. 7(c), whereas the microstructure width is nearly $20\ \mu m$. The average transit time of the particle across the microstructure is only $2\ ms$. The kinetic arrest of microspheres having such high momentum has not been previously reported. Previous reports on the kinetic arrest of microspheres were shown using holographic tweezers where the velocity of the microsphere is of the order of $\sim 100~\mu m/s$ with a trapping power of 150 $\mu W$.[19]

**Explanation of trapping/pinning at a high flow rate**

The flow in a microfluidic channel follows a laminar profile even at a high flow rate of 50 $\mu l/min$ due to low Reynolds number of the channel, $R_c = \rho u l/\mu \approx 2$ ($\rho = 1~g/cm^3$ is the density, $u = 0.016~m/s$ is the average velocity, $\mu = 0.001~kg \cdot m^{-1} \cdot s^{-1}$ is the dynamic viscosity and $l = 100~\mu m$ is the height of the channel). Microsphere inertia while moving past the microstructure can also be ignored since the lag time of microsphere $\tau_p = 2r^2\rho_p/9\mu \approx 2\mu s$ ($r = 5~\mu m$ is the radius, $\rho_p = 1.05~gm/cm^3$ is the density of particle) is much less than the characteristic disturbance time $a/U \sim 1~ms$ ($a = 10~\mu m$ is the length of downslope region and $U = 10~mm/s$ is the average velocity of microsphere near the crests). The ratio $\tau_p U/a \approx 0.002$ is known as the Stokes number, a low value of which indicates that the microsphere will respond almost instantaneously to the flow rate.[30] Inertial migration due to the parabolic flow profile can also be neglected since the microsphere Reynolds number $R_p = R_c r^2/l^2 \ll 1$.[31,32] We further carried out experiments using surfactant solution 0.1 % $PBST$, which did not result in any pinning. Therefore, trapping due to the formation of local vortices or any other fluid dynamical effects can be ruled out. Thus, we can convincingly argue that the pining of the microspheres is due to the non-specific binding occurring after they come sufficiently close to the substrate due to the hydrodynamic interaction. As the microsphere flows past the crest a sudden increase in channel volume results in low pressure in the downslope region, in other words the flow field decelerates and diverges bringing the microspheres closer to the substrate. The microspheres which are close enough will start experiencing hydrodynamic interaction with the glass substrate and gets pinned due to van der Waals interaction force. The prominent interaction forces between a spherical object and glass substrate are van der Waals and electrostatic double layer interaction force, collectively known as DLVO force. Van der Waals force between a spherical object and a flat surface is $F_V = -AR/6D^2$, where, $A$ is Hamaker constant and $R~\&~D$ are the radius of the microsphere and its distance from the substrate respectively. The electrostatic force between a spherical object and a flat surface of similar potential is $F_E = 2\pi R W_{flat}$, where, $R$ is the radius of the microsphere and $W_{flat} = 0.0211~(C)^{1/2} tanh^2[2\varphi(mV)/103]e^{-D/l_d}~Jm^{-2}$ is the interaction free energy, $D$ is the distance of microsphere from the substrate and $l_d$ is the screening length also known as the Debye length. The Hamaker constant is of the order of $A = 1 \times 10^{-20} \, J$[33] and the Debye length $l_d$ is of the order of $1.5 \, nm$, the calculation of the Debye length and other parameters is explained in the supplementary text. The polystyrene microsphere has a Zeta potential of $around \; -40 \, mV$ in $0.01 \, M$ PBS buffer at a pH of $7.5$[34] whereas the Zeta potential of fused silica ($amorphous \; Sio_2$) is around $-60 \, mV$ in $0.01 \, M$ KCl solution at a pH of $7.5$[35]. Hence, we have assumed an average potential of $\varphi = -50.0 \, mV$. Fig. 8 shows the distance dependence of van der Waals interaction force $F_v$, electrostatic double layer force $F_E$ and the total force $F_T$ for a microsphere of radius $R = 5 \, \mu m$.

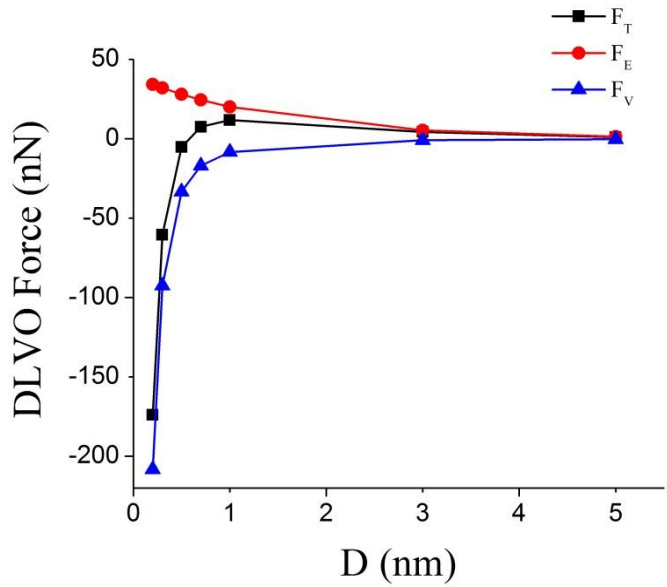

Figure 8 The van der Waals force $F_V$, the electrostatic force $F_E$ and the total DLVO force $F_T$ as a function of distance from the substrate for a microsphere of radius $5 \, \mu m$.

As the distance between microsphere and the substrate decreases the attractive van der Waals force increases much faster than the repulsive electrostatic force which results in the pinning. The van der Waals force dominates in sub-nanometre length scale which is the reason of the robust pinning of the microsphere. The distance between two surface in contact is $\sim 0.2 \, nm$[33] the respective $F_v = -206.25 \, nN$ which is much larger than the drag force generated by the flow in the pinning regime that is below $1 \, nN$ in all cases. The calculated van der Waals force $F_v = -206.25 \, nN$ is of the similar order as experimentally measured using AFM.[26] However, in the experiments which we performed, all the pinned microsphere detach at a flow rate of $1 - 2 \, ml/min$ which corresponds to a drag force of $15 - 30 \, nN$. This decrease is because, in a bead detachment experiment the microspheres are peeled off[27] rather than the transverse detachment as in the case of AFM based techniques[26]. Furthermore, we have also noticed the switching of pinned polystyrene microspheres from upslope region to downslope region. Initially, all the microspheres shown in the video "supple_pinning3.mov" were pinned in the upslope region of the microstructure ($h = 4.1\,\mu m$, $w = 5.4\,\mu m$) at a low flow rate of $0.1\,\mu l/min$. As the flow rate is increased to $30\,\mu l/min$, they switch to downslope region. Hence, the hydrodynamic interaction has an important role in facilitating robust pinning in the downslope region. In summary, the pinning on the upslope region occurs at a low flow rate because upslope acts as a barrier to particle flow whereas, pinning on the downslope region at a high flow rate is facilitated by the hydrodynamic interaction between the microsphere and substrate.

## Conclusion

We have presented a technique where near-field effects due to fluid flow past a microstructure in a channel assist the trapping and pinning of microspheres in a predictable manner. This technique enables a high-throughput production of pinned particles in any desired 2D pattern such as the chequered pattern shown in supplementary Fig. S3. However, unlike the optical tweezer based method, the trapping pattern cannot be made dynamic in this technique. This method works for the polystyrene microspheres larger than $5\,\mu m$, smaller microspheres can't be reliably pinned due to the thermal fluctuations. Nevertheless, the simple approach presented here enables the creation of a large number of pinned microspheres useful in the study of disorder in the dynamics of multi-particle interacting systems. The result from the experiment and simulation suggests that the trapping/pinning of microsphere is due to the combined effect of fluid dynamics and DLVO adhesion forces. A more quantitative analysis of many body colloidal hydrodynamics is indeed required to understand the hydrodynamic interactions. Apart from fundamental studies, pinned microspheres can also be used for molecular sensing as shown in the recent work describing a biosensor with attomolar sensitivity using a configuration of chemically patterned microspheres for detection.[36]

## Author contributions

P.P. and M.V. designed the research, analysed the data and wrote the manuscript. P.P. performed the experiments.

## Competing financial interests

The authors declare no competing financial interests.

## Data availability

Data and videos are available on the request from the corresponding author.

# Supplementary Information

**Trapping/Pinning of colloidal microspheres over glass substrate using surface features**

Praneet Prakash[1], Manoj Varma[1,2]

[1]Center for Nano Science and Engneering, Indian Institute of Science, Bangalore

[2]Robert Bosch Center for Cyber Physical Systems, Indian Institute of Science, Bangalore

[*]mvarma@cense.iisc.ernet.in

**Explanation of the real time video shown in the movie "$supple\_pinning1.mov$".**

1. ($t = 0 - 31\ sec$): Pinning of 10 $\mu m$ microspheres shown in Fig. 5(a) near the microstructure of $h = 1.1\ \mu m$, $w = 5.1\ \mu m$ (Fig. 3(a)). Initial flow rate is 50 $\mu l/min$, however, pinning starts only when the flow rate is reduced to 10 $\mu l/min$.

2. ($t = 32 - 43\ sec$): Dark field video of 5 $\mu m$ pinned microspheres shown in Fig. 5(b) near the microstructure of $h = 1.1\ \mu m$, $w = 5.1\ \mu m$ (Fig. 3(a)).

3. ($t = 44 - 1:44\ sec$): Trapping/pinning of 10 $\mu m$ microspheres over the downslope of the microstructure of $h = 4.3\ \mu m$, $w = 19.2\ \mu m$ (Fig. 3(c)) at a high flow rate of 50 $\mu l/min$ which is shown in Fig. 7(a) & 7(b).

4. ($t = 1:44 - 2:14\ sec$): Pinning is not observed at a high flow rate of 50 $\mu l/min$ on the microstructure of $h = 4.1\ \mu m$, $w = 5.4\ \mu m$ (Fig. 3(b)) unless the pump is stopped.

**Explanation of the real time video shown in the movie "$supple\_pinning2.mov$".**

1. ($t = 0 - 15\ sec$): Pinning in the upslope region at a low flow rate of 0.1 $\mu l/min$ on the microstructure of $h = 1.1\ \mu m$, $w = 5.1\ \mu m$ (Fig. 3(a)).

2. ($t = 15 - 32\ sec$): Pinning in the downslope region at a low flow rate of 0.1 $\mu l/min$ on the microstructure of $h = 4.1\ \mu m$, $w = 5.4\ \mu m$ (Fig. 3(b)).

**Explanation of the real time video shown in the movie "$supple\_pinning3.mov$".**

This video shows the switching of pinned microspheres from the upslope region to the downslope region as the flow rate is increased. Initially microspheres were pinned in the upslope region of the microstructure ($h = 4.1\ \mu m$, $w = 5.4\ \mu m$) at a low flow rate of 0.1 $\mu l/min$, as the flow rate is increased to 30 $\mu l/min$ they switch to downslope region. Hence, the hydrodynamic interaction has an important role in facilitating robust pinning in the downslope region at higher flow rates.

**Comparison of drag force calculated from simulation and analytical formula**

The drag force mentioned in the manuscript corresponds to the steady state solution, where microspheres are assumed to be pinned and a no-slip boundary condition is applied on the microfluidic channel as well as the microsphere. The accuracy of the calculated drag force by simulation can be confirmed by assuming the microsphere at the centre which will result in a drag force $F_d = 6\pi \eta r v$, where, $\eta, r$ are viscosity, radius of the sphere and $v$ is the velocity of the approaching fluid far from the microsphere. At a flow rate of $60\ \mu l/min$ analytically calculated drag force $F_{d(analytical)} = 3.09\ nN$ which is very close to the drag force calculated from the simulation $F_{d(simulation)} = 3.12\ nN$. Further, the estimation of the drag force at the bottom of the microfluidic channel is also in accordance with the analytical expression used in previous reports $F_{d(bottom)} = 1.7 \times 6\pi \eta r v$ (Reference: 22). The analytically calculated drag force at the bottom of the channel (flow rate of $60\ \mu l/min$) for a microsphere of radius $5\ \mu m$ is $F_{d(analytical)} = 0.977\ nN$ which is very close to the drag force estimated from the simulation $F_{d(simulation)} = 0.86\ nN$.

**Estimation of DLVO force between microspheres and the glass substrate**

The van der Waals force between a spherical object and a flat surface is $F_V = -AR/6D^2$, where, $A$ is Hamaker constant and $R$ & $D$ are the radius of microsphere and its distance from the substrate respectively. We have used commonly available glass slide as a substrate which contains sodium and calcium silicates ($Na_2Sio_3, Ca_2Sio_4$) along with fused silica ($amorphous\ Sio_2$). We couldn't find the Hamaker constant of plystyrene-water-glass system, however, for polystyrene-water-polystyrene system it comes out to be $0.95 - 1.3 \times 10^{-20}\ J$. Hamaker constant for other silica based solids such as fused quartz and mica in water comes out to be $\sim 0.5 - 3 \times 10^{-20}\ J$ (Ref. 33 – Intermolecular and Surface Forces, 3rd Edition, Jacob N. Israelchvili, Page No. 266). Considering the value of Hamaker constant to be $1 \times 10^{-20}\ J$ and the radius of microsphere and its distance from the substrate as $5\ \mu m$ & $0.2\ nm$ respectively, the van der Waals force comes out to be $-206.25\ nN$. Typically the distance between two surface in contact is ~ 0.2 nm (Ref. 33 – Intermolecular and Surface Forces, 3rd Edition, Jacob N. Israelchvili, Page No. 254), the respective $F_v = -206.25\ nN$ is much larger than the fluidic force generated by the flow in the pinning regime which is below $1\ nN$. The calculation of the electrostatic force $F_E$ requires the estimation of the Zeta potential of the polystyrene microsphere and glass substrate. The polystyrene microsphere has a Zeta potential of $\sim -40\ mV$ in $0.01\ M$ PBS buffer at a pH of 7.5 (Ref. 34). We were not able to find the Zeta potential of glass substrate in PBS buffer, hence, we have used Zeta potential of fused silica ($amorphous\ Sio_2$) $\sim -60\ mV$ in $0.01\ M$ KCl solution at a pH of 7.5 (Ref. 35). The range of the electrostatic force depends upon the screening of charges by the ions present in the medium which is defined by the Debye length. The Debye length $l_D$ for a low potential surface (below $\sim 25\ mV$) in a 2:2 electrolyte system such as $MgSo_4$ is defined by the expression $0.152/\sqrt{(C)}\ nm$, where $C$ is the concentration of salt solution in $Molar$ (Ref. 33 – Intermolecular and Surface Forces, 3rd Edition, Jacob N. Israelchvili, Page No. 312). We have done all our experiment in $0.01\ M$ PBS buffer, the major constituent of the buffer being $Na_2HPO_4$, which is a 3:3 electrolyte. For an order of calculation estimate, we find the Debye length by using $0.152/\sqrt{(C)}$ nm which comes out to be $1.5\ nm$, for $C = 0.01\ M$. The electrostatic force between a microsphere and a flat surface of similar potential is $F_E = 2\pi R W_{flat}$, where, $R$ is the radius of microsphere and the interaction free energy $W_{flat} = 0.0211\ (C)^{1/2} tanh^2[2\varphi(mV)/103]e^{-D/l_d}\ Jm^{-2}$. The electrostatic force (repulsive) $F_E$ at a contact distance of $0.2\ nm$ (Ref. 33 – Intermolecular and Surface Forces, 3$^{rd}$ Edition, Jacob N. Israelchvili, Page No. 254), is $34.3\ nN$ whereas van der Waals force (attractive) is $-206.25\ nN$, hence, the van der Waals force dominates in the sub-nanometre length scale.

**Table 1** Parameters for wet etching.

| Mirostructure | Photoresist | Resist width | Periodicity | Post-bake time | HF etch duration |
|---|---|---|---|---|---|
| $h = 1.1\ \mu m, w = 5.1\ \mu m$ | S1813 | $5\ \mu m$ | $50\ \mu m$ | 3 min @ 110°C | $10 - 15\ sec$ |
| $h = 4.1\ \mu m, w = 5.4\ \mu m$ | AZ4562 | $5\ \mu m$ | $50\ \mu m$ | 5 min @ 110°C | $40 - 45\ sec$ |
| $h = 4.3\ \mu m, w = 19.2\ \mu m$ | AZ4562 | $20\ \mu m$ | $100\ \mu m$ | 5 min @ 110°C | $40 - 45\ sec$ |

**Figure S1**

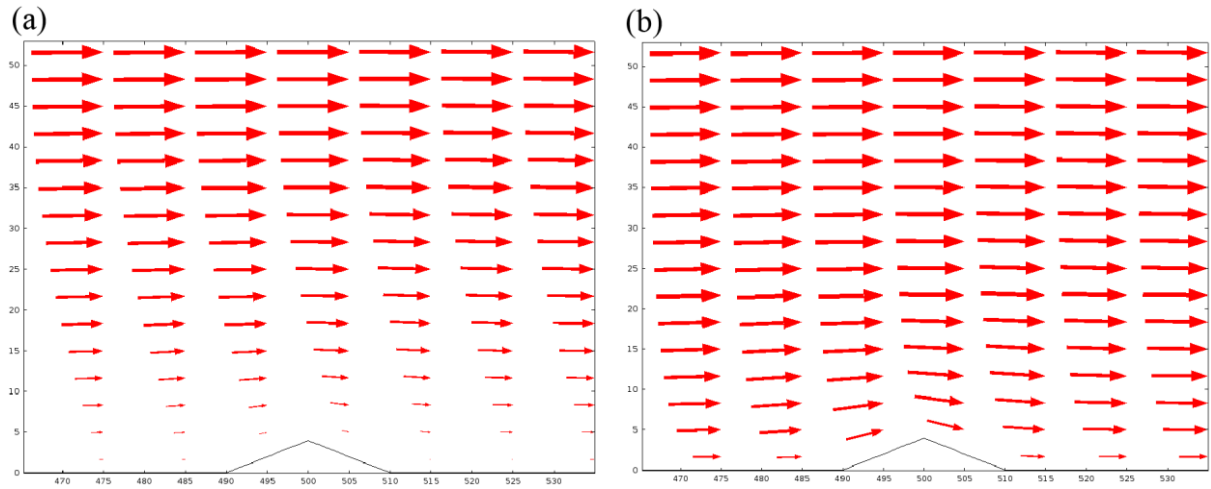

**Fig. S1** Velocity vector field over the surface of microstructure ($h = 4\ \mu m,\ w = 20\ \mu m$) at a volumetric flow rate of $50\mu l/min$. Here we have shown only half of the channel which is from the bottom to the middle. (a) Arrow of the velocity vector field is proportional to the magnitude of the velocity with a maximum velocity of $2.5\ cm/s$ at the top of the graph. (b) Arrow of the velocity vector field is proportional to the logarithmic of the magnitude of velocity so that the velocity vector arrows which are very near to the bottom surface are also visible.

**Figure S2**

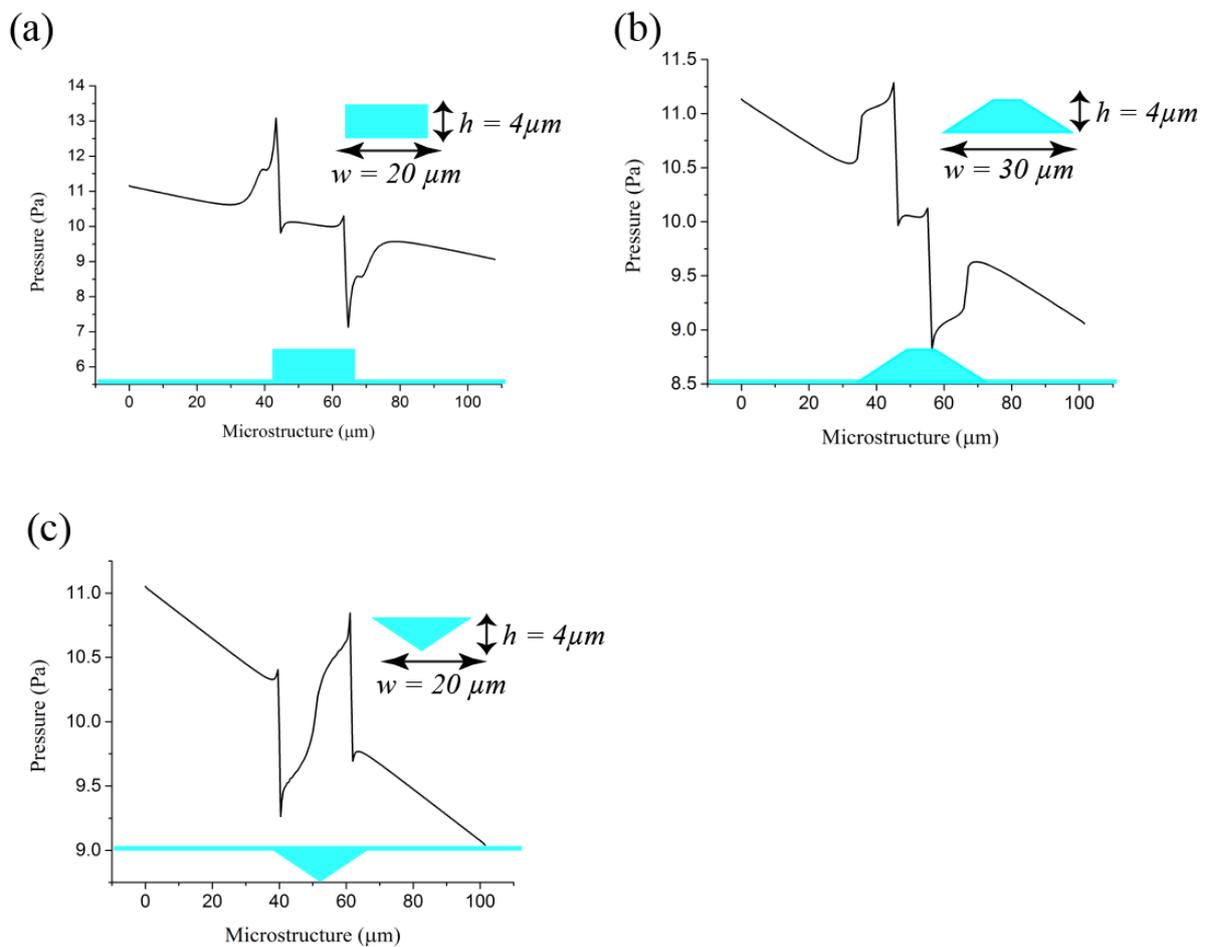

**Fig. S2** Pressure profile plots over the surface of microstructure with different geometries. Volumetric flow rate was $50 \mu l/min$ and the channel domain was taken to be $1\ mm$ long. (a) Square (b) Trapezoid (c) Triangular shaped trough.

Fabrication of a square shaped crests isn't possible with isotropic wet etch, however, Deep reactive-ion etching can be employed to fabricate them. Trough structures can be explored to see their pinning ability, however, microspheres should be very close to the substrate surface for it to interact with the lower regions and hence, random pinning may increase.

**Figure S3**

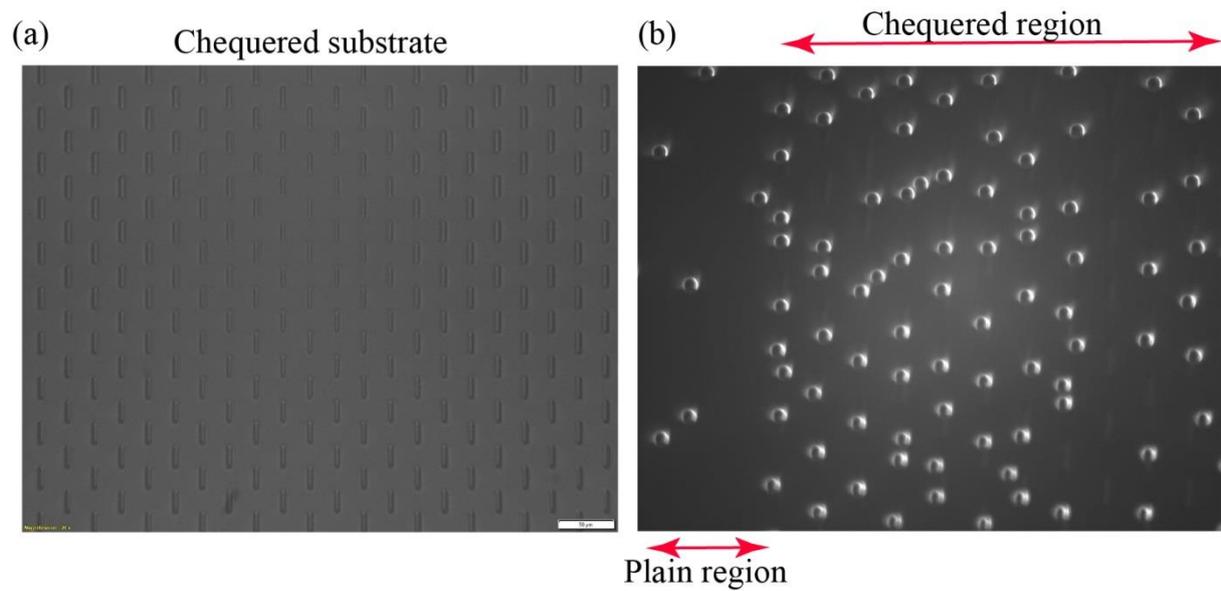

**Fig. S3** Pinning over a chequered substrate. (a) A chequered substrate with alternate microstructures of $h = 1.1\ \mu m,\ w = 6.5\ \mu m$. (b) Chequered pattern of microspheres enabled by the positioning of microstructures.

A substrate such as shown above can enable the precise positioning of microspheres according to the pattern of microstructure.